\documentclass[aps,pre,reprint,superscriptaddress,showpacs,amsmath,amssymb]{revtex4-1}
\usepackage{graphicx}

\newcommand{\Eq}[1]{Eq.~(\ref{#1})}
\newcommand{\Fig}[1]{Fig.~(\ref{#1})}
\newcommand{\Dif}[0]{d}

\begin{document}


\title{Self-similar dynamics of bacterial chemotaxis}


\author{Waipot Ngamsaad}
\email{waipot.ng@up.ac.th}
\affiliation{Division of Physics, School of Science, University of Phayao, Mueang Phayao, Phayao 56000, Thailand}

\author{Kannika Khompurngson}
\affiliation{Division of Mathematics, School of Science, University of Phayao, Mueang Phayao, Phayao 56000, Thailand}
\affiliation{Centre of Excellence in Mathematics, PERDO, CHE, Thailand}


\date{\today}

\begin{abstract}
Colonies of bacteria grown on thin agar plate exhibit fractal patterns as a result of adaptation to their environments. The bacterial colony pattern formation is regulated crucially by chemotaxis, the movement of cells along a chemical concentration gradient. Here, the dynamics of pattern formation in bacterial colony is investigated theoretically through a continuum model that considers chemotaxis. In the case of the gradient sensed by the bacterium is nearly uniform, the bacterial colony patterns are self-similar, which they look the same at every scale. The scaling law of the bacterial colony growth has been revealed explicitly. Chemotaxis biases the movement of bacterial population in colony trend toward the chemical attractant. Moreover, the bacterial colonies evolve long time as the traveling wave with sharp front. 
\end{abstract}


\maketitle

Bacteria adapt to the hostile environmental conditions by cooperatively spreading the colony with well-defined spatial patterns \cite{Ben-Jacob2000, Murray2002}. The colonies of various bacterial species exhibit branching pattern \cite{Matsuyama1992, Kawasaki1997, Golding1998, Ben-Jacob2000} that looks similar to the fractal pattern in the diffusion-limited aggregation (DLA) process \cite{Witten1981}. In this manner, the branching patterns of bacterial colony are typically self-similar, where they are the same at every scale (scale invariant). It suggests that the pattern formation reflects the bacterial communication and social behavior \cite{Ben-Jacob2004, Ben-Jacob2009}. The underlying mechanism of the bacterial pattern formation is important because it is a key to understand living organisms. 

The bacteria respond to a chemical attractant such as nutrient by swimming along its gradient, known as the chemotaxis. It has demonstrated that the chemotaxis has an essential role on the regulation of bacterial colony pattern formation \cite{Adler1966, Budrene1991, Budrene1995}. The bacteria move in fluid medium by swimming as random walk motion, in which the bacteria propel themselves in nearly straight run separated by brief tumble to change directions. They detect the spatial gradients by comparing a temporal difference between the amounts of attractant molecules that bind to the membrane receptors along their path. Then the bacteria delay the tumbling frequency as cells swim up the gradient of the attractant (or down the gradient of repellent). This causes the bacteria move in directions of increasing attractant gradient.

The reaction-diffusion model has successfully described the dynamics of pattern formation in bacterial colonies at continuum level \cite{Kawasaki1997, Golding1998, Ben-Jacob2000, Murray2002}. Recently, a nonlinear reaction-diffusion with chemotaxis model has been proposed for studying the pattern formation in bacterial colonies exemplified by \textit{Paenibacillus dendritiformis} grown on Petri dish \cite{Golding1998, Cohen1999, Kozlovsky1999, Ben-Jacob2000}. This bacteria species is motile on the dry surface by cooperatively producing a layer of lubrication fluid in which they swim. Its colony exhibits branching pattern. The numerical simulations of this model can reproduce the branching pattern in the bacterial colonies, which agrees well in comparison with experimental data. However, the scaling law that indicates the self-similarity of bacterial colony growth has not been obtained explicitly by the numerical results. Therefore, an analytical work is needed to be carried out. 

In this work, we investigate the simplified form of nonlinear reaction-diffusion with chemotaxis models for pattern formation in bacterial colony \cite{Golding1998, Cohen1999, Kozlovsky1999, Ben-Jacob2000}. The aim of this paper is to find the scaling law of bacterial colony evolution. This analytical result could be plausible for interpreting the results in both experiments and simulations. 

We now explain the bacterial chemotaxis model in our consideration. The bacterial colony evolves in two dimensions however; each tip grows in one dimension, except for occasional branching. This allows us to investigate this problem in one-dimensional space, which its results could be equivalent to one obtained from the two-dimensional space \cite{Kawasaki1997}. As proposed in Ref. \cite{Gilding2005}, the dynamics of bacterial populations is governed by a generalized convection-reaction-diffusion equation 
\begin{equation}\label{eq:gen_RD}
\frac{\partial u}{\partial t} = \frac{\partial }{\partial x}\left(D(u)\frac{\partial u}{\partial x} - u\vartheta(s) \right) + R(u) , 
\end{equation} 
where $u(x,t)$ and $s(x,t)$ are, respectively, the bacterial density and the attractant density in spatial coordinate $x$ and time $t$. $D(u)$, $R(u)$ and $\vartheta(s)$ are the diffusion coefficient, the reaction term and the drift velocity due to the chemotaxis, respectively. This equation is similar to the generalized Keller-Segel equation \cite{Tindall2008}. \Eq{eq:gen_RD} is a simplified form of the full model in Refs. \cite{Golding1998, Cohen1999, Kozlovsky1999}, under following assumptions. i) The nutrient density is proportional to the bacterial density and it is absorbed some way in reaction term. ii) The production of lubricant fluid is proportional to the bacterial density and it is absorbed into the medium. Thus, the effect of lubrication fluid is represented through the diffusion \cite{Cohen1999, Gilding2005}. iii) The chemotactic signal can be also a field produced directly or indirectly by the bacterial cells. 

We consider the diffusion coefficient in a density-dependent form $D(u) = M (u/\sigma)^p$, where $M > 0$ is diffusion constant, $\sigma = \lim_{t \to \infty} u(x,t)$ is equilibrium density and $p > 0$. This represents the crowd-avoidance movement of individuals \cite{Gurney1975, Gurney1976, Gurtin1977, Newman1980, Newman1983}. The growth with limited nutrient supply of bacteria is modeled as the generalized logistic law $R(u) = \alpha u[1-(u/\sigma)^p]$, where $\alpha > 0$ is rate constant \cite{Newman1983}. The chemotatic drift velocity can be expressed as $\vartheta(s) = \zeta(s)\chi(s)s_x$, where $\chi(s)s_x$ acts as the gradient sensed by the bacterium (with $\chi(s)$ having the units of 1 over the chemical concentration) \cite{Cohen1999}. $\zeta(s)$ is the bacterial response to the sensed gradient and it has the same units as a diffusion coefficient \cite{Cohen1999}. Therefore, we assume that $\zeta(s) = \gamma D(u) = \gamma M (u/\sigma)^p$, where $\gamma$ is a constant, positive for attractive chemotaxis and negative for repulsive chemotaxis \cite{Cohen1999}. Here, we are interested in a special case where the gradient sensed by the bacterium is nearly uniform and $\chi(s)s_{x}$ is treated as a constant \cite{Gilding2005}. By substituting $D(u)$, $R(u)$ and $\vartheta(s)$ into \Eq{eq:gen_RD} with the transformations $t^\ast = \alpha t$, $x^\ast = (m \alpha /M)^\frac{1}{2}x$, $u^\ast = u/\sigma$ and $\kappa = (1/2)\gamma  (m M / \alpha)^\frac{1}{2}\chi(s)s_{x}$, we obtain the dimensionless equation
\begin{equation}\label{eq:convec_1}
u_t  = (u^{m})_{xx} - 2\kappa(u^{m})_x +u -u^{m} ,
\end{equation} 
where $m = p+1 > 1$ and the asterisk is dropped. So far, the solution of \Eq{eq:convec_1} has well understood as the traveling wave \cite{Rosenau2002, Gilding2005, Mansour2010}. However, the exact or explicit solution in space-time coordinates has been unknown.

As studied in our previous work, without chemotaxis, \Eq{eq:convec_1} can be mapped to a purely diffusion process, which the exact solution can be obtained \cite{Ngamsaad2012}. We then extend the similar technique to analyze \Eq{eq:convec_1}. We rewrite \Eq{eq:convec_1} as $\left(\frac{\partial}{\partial t} - 1\right)u = \frac{1}{\omega^{2}}\left(\frac{\partial}{\partial y} +1\right)\left(\frac{\partial}{\partial y} - \omega^2\right)u^m$, where $y = x/\omega$ and $\omega = \kappa \pm \sqrt{\kappa^2+1}$, and then it can be evaluated to $e^{t}\frac{\partial}{\partial t}e^{-t} u = \omega^{-2} e^{-y}\frac{\partial}{\partial y}e^{y}(e^{\omega^2 y}\frac{\partial}{\partial y}e^{-\omega^2 y} u^m)$. By introducing the transformations 
\begin{eqnarray}\label{eq:main_sol}
u(y,t) &=& e^{t} e^{\frac{\omega^2}{m} y} \Phi(y,t)  \\
\tau(t) &=& e^{(m-1)t} - 1 \\
\phi(y) &=& e^{ \frac{(m+\omega^2)}{m} y } ,
\end{eqnarray} 
we obtain the reduced form of \Eq{eq:convec_1}
\begin{equation}\label{eq:tran_convec_1}
\Phi_{\tau} = k[\phi^{l} (\Phi^m )_{\phi}]_{\phi} ,
\end{equation}
where $k = \frac{1}{\omega^2(m-1)}\left(\frac{m+\omega^2}{m}\right)^2$ and $l = \frac{(\omega^2+2)m+\omega^2}{m+\omega^2}$. \Eq{eq:tran_convec_1} is known as the anomalous diffusion equation, whose solution is assumed to be the scaling function $\Phi(\phi,\tau) = \frac{1}{T(\tau)}F\left(\frac{\phi}{T}\right) = \frac{F(\theta)}{T(\tau)}$, where $\theta(\phi,\tau) = \phi/T(\tau)$ \cite{Bologna2000, Tsallis2002, Lenzi2003}. By performing the calculations similar to Ref. \cite{Ngamsaad2012}, we obtain
\begin{equation}\label{eq:self-sim_convec_1}
\Phi = \left\lbrace \frac{1}{(\tau + a)^\frac{m+\omega^2}{m}}\left[ b + \theta^{-\frac{(m-1)\omega^2}{m+\omega^2}} \right]\right\rbrace^\frac{1}{m-1} ,
\end{equation}
where $a>0$ and $b$ are constant. After substituting \Eq{eq:self-sim_convec_1} into \Eq{eq:main_sol}, we obtain the initial density profile: $u_0(x) = u(x,0) = a^{-1/p} \left\lbrace  1 +b\left[ e^{p\omega x}/a^{\omega^2} \right]^{1/p+1}  \right\rbrace^{1/p}$. We consider the initial density that satisfies the following properties: $u_0(x)=0$ for $x\geqslant x_0$ and $\lim_{x \to -\infty}u_0(x)=\rho$, where $\rho$ is initial density amplitude and $x_0$ is initial front position \cite{Ngamsaad2012}. According to these conditions, we have $a = \rho^{-p}$ and $b = - \rho^{-p\omega^2/(p+1)} e^{-p\omega x_{0}/(p+1)}$. Now the exact solution to \Eq{eq:convec_1} is given by
\begin{eqnarray}\label{eq:convec_sol_exact}
\lefteqn{ 
u(x,t) = \frac{ \rho e^{t} }{ [\rho^{p}( e^{pt} -1) +1]^\frac{1}{p} } 
} \nonumber\\ && \times
\left\lbrace  1 - \left[ \frac{ e^{p\omega(x-x_{0})} }{ [\rho^{p}( e^{pt} -1 ) + 1]^{\omega^2} } \right]^\frac{1}{p+1}  \right\rbrace   ^ \frac{1}{p} .
\end{eqnarray}

Since the solution \Eq{eq:convec_sol_exact} has two forms, depending on the value of $\omega$, we define $u_{+}(x,t)$ and $u_{-}(x,t)$ as the solutions corresponding to $\omega_{+}=\kappa + \sqrt{\kappa^2+1}$ and $\omega_{-}=\kappa - \sqrt{\kappa^2+1}$, respectively. As proved in our previous work \cite{Ngamsaad2012}, the linear combination of these two solutions $w(x,t) = u_+(x,t)+u_{-}(x,t)$ is a solution of \Eq{eq:convec_1}. By using an approximation $(u_++u_-)^p \approx u_+^{p}+u_-^{p}$ \cite{Ngamsaad2012}, we obtain 
\begin{equation}
\label{eq:main_sol_exact_sym}
w(x,t) \approx  2^{-\frac{1}{p}}\left[ u_+^p(x,t)+u_-^p(x,t) \right]^{\frac{1}{p}} ,
\end{equation}
where $(2)^{-1/p}$ is normalized factor. We note that in the case of no chemotaxis $\kappa=0$, thus $\omega = \pm 1$, these results recover our previous work \cite{Ngamsaad2012}.  

\begin{figure}[htp]
\center\includegraphics[width=\columnwidth]{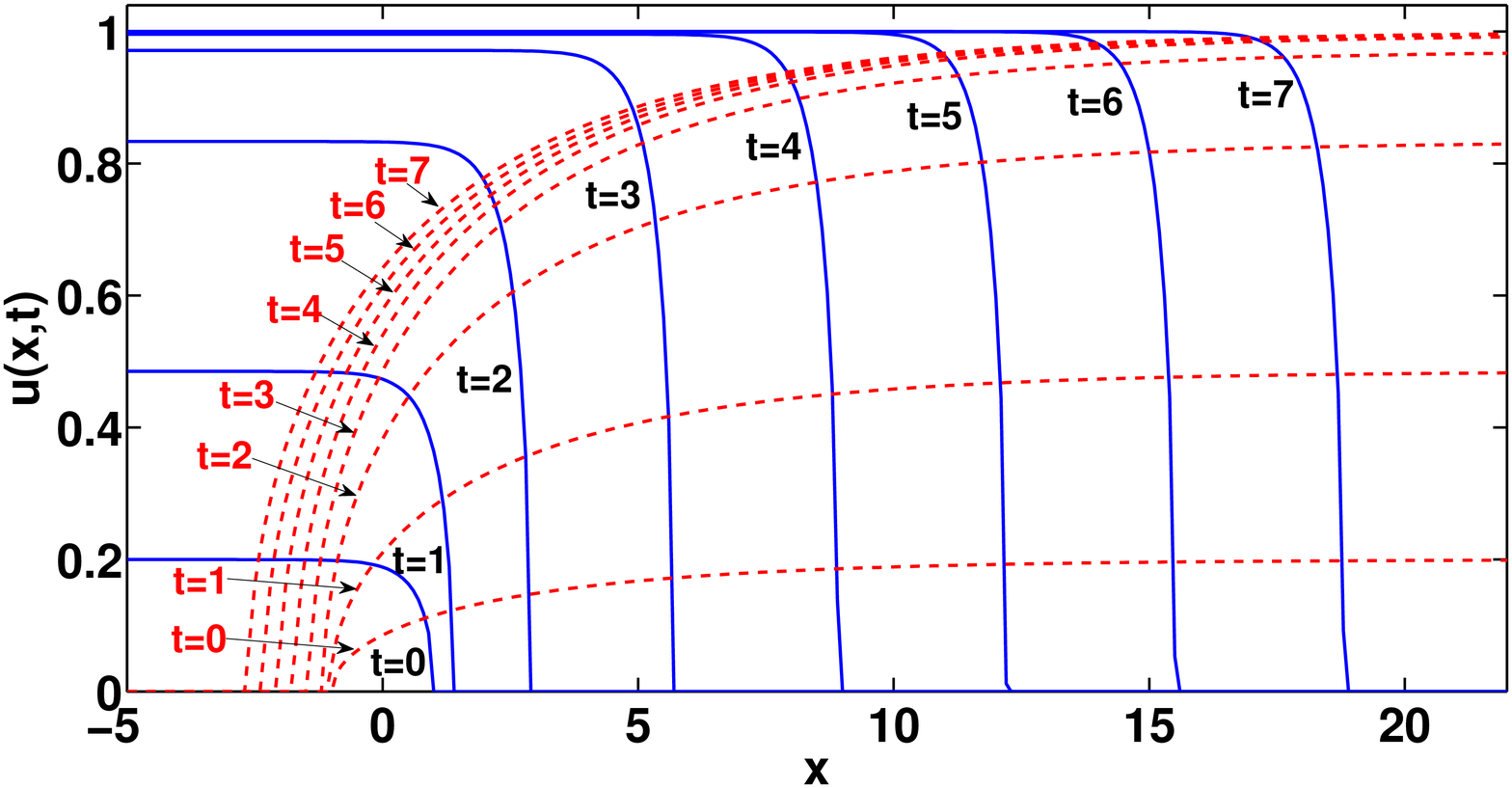}
\caption{\label{fig:PDF_convec}
(Color online) The spatiotemporal evolution of the bacterial density profile $u(x,t)$ (\Eq{eq:convec_sol_exact}) in the case of $p=2$ with initial conditions $\rho=0.2$ and $x_{0}=1$. The solid lines represent $u_+(x,t)$ and the dashed lines represent $u_-(x,t)$.
}
\end{figure}

\begin{figure}[htp]
\center\includegraphics[width=\columnwidth]{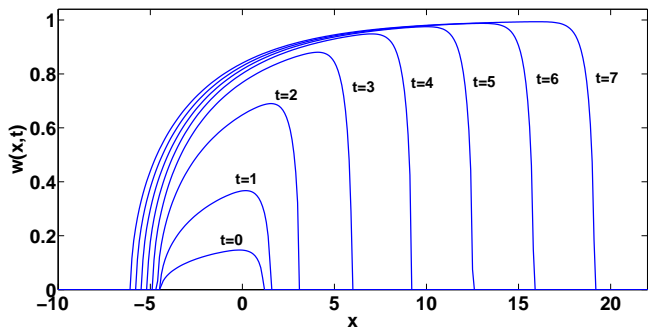}
\caption{\label{fig:PDF_convec_sym}
(Color online) The spatiotemporal evolution of the pulse-like bacterial density profile $w(x,t)$ (\Eq{eq:main_sol_exact_sym}) in the case of $p=2$ with initial conditions $\rho=0.2$ and $x_{0}=1$.
}
\end{figure}

The evolution in space and time of bacterial density profiles $u_{+}(x,t)$ and $u_{-}(x,t)$, as in \Eq{eq:convec_sol_exact}, is illustrated in \Fig{fig:PDF_convec}. The density profiles start from the initial state $u_0(x)$ then grow and expand to the unoccupied region. At sufficient large time scale, the density profiles reach the saturated value at 1. After that, they seem to propagate with unchanged shape; $u_+(x,t)$ is propagating to the right whereas $u_{-}(x,t)$ is propagating to the left. The roles of chemotaxis on the regulation of pattern formation in the system is reflected by parameter $\kappa$. Since $\kappa < \sqrt{\kappa^2+1}$, $\omega_{+}$ is always positive whereas $\omega_{-}$ is always negative. This causes the tails of $u_+$ decay as $x\to\infty$ and the tails of $u_-$ decay as $x\to -\infty$. In the former case, the front-interface is sharper because $|\omega_{+}| > |\omega_{-}|$. Due to the influence of chemotaxis, the distribution of density profile is biased toward to the right thus the front of $u_+$ is moving faster than of $u_-$. The spatiotemporal evolution of the combined density profiles $w(x,t)$ is also illustrated in \Fig{fig:PDF_convec_sym}. The densities $w(x,t)$ form the pulse-like profiles that grow and expand with asymmetric shape. It behaves like $u_+(x,t)$ for $x \gg x_0$ and like $u_-(x,t)$ for $x \ll -x_0$. Due to the bias force from chemotaxis, the peak of $w(x,t)$ is moving toward to the right.

\begin{figure}[htp]
\center\includegraphics[width=\columnwidth]{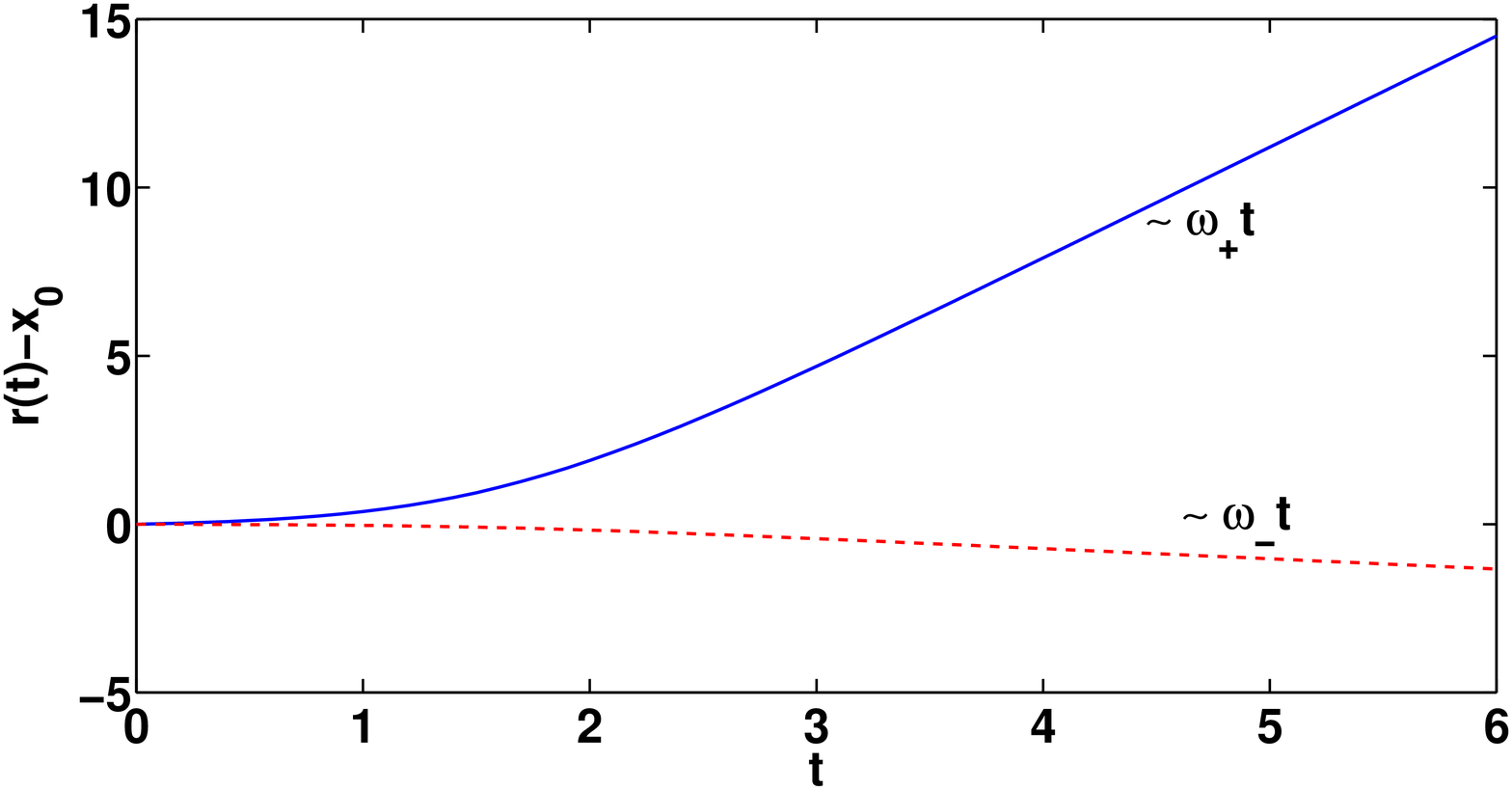}
\caption{\label{fig:Front_convec}
(Color online) The relative front position $r(t)-x_0$ corresponding to the bacterial density profile in \Fig{fig:PDF_convec}. The solid line represents the relative front position of $u_+(x,t)$ and the dashed line represents the relative front position of $u_-(x,t)$.
}
\end{figure}

From \Eq{eq:convec_sol_exact}, we calculate the front position $r(t)$, that the density falls to zero $u(r,t) = 0$ as $r(t) = x_{0} + \omega\frac{\ln [ \rho^p(e^{pt} -1) + 1 ] }{p}$. The plot of relative front position $r(t) - x_{0}$ is shown in \Fig{fig:Front_convec}. The relative front position of $u_-(x,t)$ is slow varying when compared with of $u_+(x,t)$. At sufficient large time, that $e^{pt^\prime}\gg 1$ and $\rho^p e^{pt^\prime} \gg 1$ thus $t^{\prime} \approx  -\ln \rho$, the relative front position seems to vary linearly in time $r(t) - x_{0} \sim \omega t$. It implies the constant front propagating speed.  Consequently, we calculate the front speed as $v(t) = \frac{\Dif}{\Dif t}r(t) = \frac{\omega\rho^p e^{pt} }{ \rho^p(e^{pt} -1) + 1 }$. At large time scale $t \gg t^\prime$, the front speed trends to be constant $c=\lim_{t\to\infty}v(t) = \omega(\kappa)$. At this point, it is clearly seen that the spreading speed is biased by the chemotaxis through the parameter $\kappa$.

At the large time scale, $t\gg t^{\prime}$, the bacterial density profile \Eq{eq:convec_sol_exact} emerges the traveling wave form
\begin{equation}\label{eq:convec_travel_wave_1}
\widetilde{u}(x-\omega t) = \left[ 1 - \frac{e^{\frac{p\omega}{p+1}(x - \omega t - x_{0})}}{\rho^{\frac{p\omega^2}{p+1}}}  \right] ^ \frac{1}{p} ,
\end{equation}
where $\omega$ is front speed. The front speed obtained here is comparable to the minimum value for the sharp traveling wave \cite{Gilding2005, Mansour2010}. Similarly, at the large time scale $t \gg t^{\prime}$, \Eq{eq:main_sol_exact_sym} develops to the expanding pulse-like wave 
\begin{equation}
\label{eq:main_sol_exact_sym_wave}
\widetilde{w}(x-\omega_{\pm}t) \approx  2^{-\frac{1}{p}}\left[ \widetilde{u}_+^p(x-\omega_+ t)+\widetilde{u}_-^p(x-\omega_- t) \right]^{\frac{1}{p}} ,
\end{equation}
with the expanding speed $\omega_{\pm}$. 

Finally, we found that \Eq{eq:self-sim_convec_1} forms a scaling law at large time scale $t \gg t^{\prime}$
\begin{equation}\label{eq:scaling_law}
\Phi(\phi,\tau) \approx \frac{1}{\tau^\beta} F(\frac{\phi}{\tau^\beta}) , 
\end{equation}
where $\beta = \frac{m+\omega^2}{m(m-1)}$. It implies that the bacterial colonies evolve as the self-similar object in the terms of transformed quantities: $\Phi \to e^{-\omega x/m} e^{-t} u$, $\tau \to e^{(m-1)t}$, and $\phi \to e^{(m+\omega^2)(x-x_0)/m\omega}$. Moreover, they evolve from  self-similar pattern form to the traveling wave pattern form.  This behavior can be classified as the intermediate asymptotics of the second type \cite{Barenblatt1972}.

In summary, the spatiotemporal pattern formation of bacterial colony in the presence of chemotaxis has been investigated at continuum level. We have shown that the bacterial colony patterns in the case of uniform gradient sensed by bacterium are self-similar; where they are scale invariant. The scaling law of bacterial colony growth has been revealed explicitly. Moreover, we found that the bacterial colonies evolve long time scale as the sharp traveling wave where the front speed is biased to move toward to the chemical attractant. 

K. Khompurngson acknowledges the Centre of Excellence in Mathematics (Thailand) for partial financial support. 

\bibliography{DDRDE_ref}


\end{document}